\documentclass[reprint,
 amsmath,amssymb,
 aps,
physrev
]{revtex4-2}

\usepackage{graphicx}% Include figure files
\usepackage{bm}% bold math
\begin{document}

\preprint{APS/123-QED}
\title{Frequency chimera state induced by differing dynamical timescales}% Force line breaks with \\

\author{Sneha Kachhara}
 %\email{snehakachhara@students.iisertirupati.ac.in}
 \affiliation{Indian Institute of Science Education and Research (IISER) Tirupati, Tirupati, 517507, India}
\author{G. Ambika}
 \email{g.ambika@iisertirupati.ac.in}
 \affiliation{Indian Institute of Science Education and Research (IISER) Tirupati, Tirupati, 517507, India}

%date{\today}

\begin{abstract}
We report the occurrence of a self-emerging frequency chimera state in spatially extended systems of coupled oscillators, where the coherence and incoherence are defined with respect to the emergent frequency of the oscillations. This is generated by the local coupling among nonlinear oscillators evolving under differing dynamical timescales starting from random initial conditions. We show how they self-organize to structured patterns with spatial domains of coherence that are in frequency synchronization, coexisting with domains that are incoherent in frequencies.  
Our study has relevance in understanding such patterns observed in real-world systems like neuronal systems, power grids, social and ecological networks, where differing dynamical timescales is natural and realistic among the interacting systems.

\end{abstract}


\maketitle

\section{Introduction}
Among the variety of the collective behavior observed in coupled nonlinear oscillators, the chimera state is the most fascinating and intriguing. It is a state in which coherence and incoherence coexist in the spatiotemporal patterns of the system \cite{parastesh2020chimeras}. Although initially limited to phase chimeras \cite{abrams2004chimera}, with nonlocal coupling and specially prepared initial conditions, subsequent studies indicate that  chimeras can occur with local \cite{laing2015chimeras} or global coupling \cite{sethia2014chimera} and distance-dependent power-law coupling \cite{banerjee2016chimera} and that may not even require special initial conditions in many cases \cite{omel2008chimera,zhang2021mechanism}. Recent research established the occurrence of chimera states in complex networks \cite{zhu2014chimera,scholl2019chimera,brezetsky2021chimera}, star networks\cite{meena2016chimera}, multilayer networks \cite{ghosh2016emergence}, in time varying topologies \cite{buscarino2015chimera}, networks with hierarchical connectivity \cite{ulonska2016chimera}, structured heterogeneous networks \cite{li2017chimera}, two dimensional (2D) lattices with fractal connectivity \cite{argyropoulos2019chimera}, and oscillating medium \cite{bolotov2018simple}. Interest in this special type of emergent behavior is sustained mainly by the fact that it is understood to have a prominent role in brain function and disease \cite{majhi2019chimera,wang2020brief} especially in the context of unihemispheric sleep of some birds and marine mammals \cite{rattenborg2000behavioral,mathews2006asynchronous} and first-night effect in human sleep \cite{tamaki2016night}. Hence this phenomenon is studied in great details in models of coupled spiking and bursting neurons \cite{bera2016chimera,majhi2019chimera,santos2017chimera}. It has been observed in many experimental set ups and physical systems like optomechanical arrays \cite{pelka2020chimera}, chemical oscillators \cite{tinsley2012chimera}, mechanical oscillators \cite{martens2013chimera}, and active camphor ribbons \cite{sharma2021chimeralike}.\\

We note that the notion of chimera state itself seems to be changing and expanding within the last decade \cite{parastesh2020chimeras,haugland2021changing} to encompass a wide range of phenomena, with several types added, like amplitude chimera \cite{verma2020amplitude,bogomolov2017mechanisms}, amplitude-mediated chimera \cite{sethia2013amplitude}, chimeralike states \cite{dutta2015spatial}, chimera death \cite{zakharova2014chimera}, phase-flip chimera \cite{chandrasekar2016phase}, breathing chimera \cite{abrams2008solvable}, traveling chimera \cite{bera2016imperfect}, coherence-resonance chimera \cite{semenova2016coherence}, multichimeras \cite{omelchenko2013nonlocal,ujjwal2013chimeras}, spiral wave chimera \cite{martens2010solvable}, self-propelled chimeras \cite{kruk2018self} etc.\\

In this article, we present our study on a new type of frequency chimera in which the coherence and incoherence are characterized in terms of the frequency of the oscillations, coherence corresponds to frequency synchronized state and incoherence relates to varying frequencies. This is generated by the coupling among nonlinear oscillators that evolve under differing dynamical timescales. \\

Natural systems evolve with many intrinsic rhythms and cycles and often different timescales occur in a single system, such as models of neurons. But in complex systems, this difference in timescales can manifest across the constituent dynamical units as well. Thus a network of oscillators with differing dynamical timescales is a better approximation to realistic interacting systems such as neuronal networks, power grids, social networks, etc. Hence a study on the possibility of emergent states in such a network is novel and relevant. We find that this type of mismatch in dynamical timescales can induce a new type of frequency chimera state, without special initial conditions or nonlocal coupling. We note that this mismatch in timescales is different from nonidentical intrinsic frequency, or mismatch in the timescale among the variables of the same intrinsic dynamics. \\

The article is organized as follows. We introduce the governing equations for coupled oscillators in section \ref{sect:model}. We present the frequency chimera state in section \ref{sect:freq_chimera}, and chimeralike and chaotic traveling wave states in section \ref{sect:chimeralike}, for coupled oscillators with R\"{o}ssler dynamics. The characterization and conditions of emergence of such states are discussed in \ref{sect:characterization}. We also discuss the emergence of frequency chimera states in coupled van der Pol oscillators, in \ref{sect:freq. chimera vdP}. Our main results with their scope and relevance are summarized in section \ref{sect:conclusion}.

\section{Governing Equations}\label{sect:model}
We start by considering the framework of a system of coupled nonlinear oscillators that in general, is described by:
\begin{equation}\label{eq:mismatch}
    \mathbf{\dot{X_{i}}} = \tau _{i}\mathbf{F(\mathbf{X_{i}})}+\tau_{i}\varepsilon\mathbf{H}(\mathbf{X_{i}},\mathbf{X_{j}})
\end{equation}
where i =  1,2...N. Here N is the number of oscillators and $\mathbf{X_{i}}$ is an n-dimensional vector representing the state of the $i^{th}$ oscillator.
$\mathbf{F}(\mathbf{X_{i})}$ represents the intrinsic dynamics of the $i^{th}$ oscillator, and $\mathbf{H}(\mathbf{X_{i},X_{j}})$ represents the coupling function between oscillators i and j. Here the dynamical timescale of each oscillator is decided by the parameter $\tau_{i}$. Without loss of generality, we take the case of two different timescales in the system by considering a mismatch parameter $\tau$ such that $\tau$ $<$ 1, would mean a slow timescale for the oscillators, relative to the fast oscillators that are assigned $\tau$ = 1. Then, decreasing the value of $\tau$ increases the mismatch between oscillators.\\

The other parameter of relevance is the coupling strength $\varepsilon$ and the emergent behavior of the coupled system depends mostly on the choice of these two parameters, $\tau$ and $\varepsilon$. For the specific case of a coupling pattern corresponding to complex networks of random or scale-free type, emergent phenomena like suppression of dynamics, cluster synchronization, and reorganization into a state of common emergent frequency are reported for coupled slow and fast systems\cite{gupta2019role}. Also, the inhibitory coupling between the neuronal modules with two timescales, is reported to generate synchronized frequency locked clusters with traveling burst sequences of recurring patterns \cite{mozumdar2019frequency}.\\

We consider the standard spatially extended system as a ring with diffusive coupling. 
As nodal dynamics we take the R\"{o}ssler oscillator, which is a prototypical nonlinear oscillator and study the possible collective behavior in a system of coupled oscillators in this framework. The dynamics is then given by, 

\begin{eqnarray}\label{eq:ross}
    \frac{\mathrm{d} x_{i}}{\mathrm{d} t} & = & \tau_{i} (-y_{i}-z_{i} )+\tau_{i} \varepsilon \sum_{j = i-P}^{i+P}(x_{j}-x_{i})\nonumber\\
    \frac{\mathrm{d} y_{i}}{\mathrm{d} t} & = & \tau_{i} (x_{i}+ay_{i} )\nonumber\\
    \frac{\mathrm{d} z_{i}}{\mathrm{d} t} & = & \tau_{i} (b+z_{i}(x_{i}-c))\nonumber\\
\end{eqnarray}
with periodic boundary conditions. Here a, b and c are parameters that are chosen such that the local dynamics is in the chaotic regime (a = b = 0.1, c = 18). P represents the range of coupling (P = 1 for nearest neighbor coupling). The system is evolved numerically with random initial conditions, with $N_s$ randomly chosen systems as slow with their $\tau_{i} < $ 1. We show that the variations in the two parameters $\tau$ and $\varepsilon$, result in the system self-adjusting to a rich variety of spatiotemporal dynamics such as frequency chimera, chimeralike states, and multicluster states, in addition to amplitude death (AD) and frequency synchronized state for all the oscillators. \\

We also study coupled van der Pol oscillators, which model relaxation oscillations in a similar configuration. \\
The governing equations in this case are:
\begin{eqnarray}\label{eq:vdp}
    \frac{\mathrm{d} x_{i}}{\mathrm{d} t} & = & \tau_{i}y+\tau_{i} \varepsilon \sum_{j = i-P}^{i+P}(x_{j}-x_{i})\nonumber\\
    \frac{\mathrm{d} y_{i}}{\mathrm{d} t} & = & \tau_i \mu(1-x^2)y-\tau_i x+\tau_iA\cos (\omega t)\nonumber\\
\end{eqnarray}
with periodic boundary conditions. Here, $\mu = 8.53$ is the nonlinear damping parameter, $A = 1.2$ is the forcing amplitude, and $\omega = \frac{2\pi}{10}$ is the angular frequency. For this set of parameter values, the intrinsic dynamics is chaotic\cite{mettin1993bifurcation}. The system is evolved with random initial conditions, with $N_s$ randomly chosen oscillators as slow with their $\tau_{i} < $ 1.\\

\section{Frequency chimera states}\label{sect:freq_chimera}

For a ring of $N = 100$ oscillators, we take $N_s = 50$ slow nodes distributed randomly over the ring. When started with initial conditions that are randomly distributed between -1 and 1, under suitably chosen values of $\varepsilon$  and $\tau$, we observe the coexistence of multiple domains with coherence in frequency, separated by domains of incoherence in frequency. We call this interesting new state a frequency chimera state. The domains with synchronized frequency do not have synchronized amplitude or phase and hence this state is different from
amplitude-mediated chimera\cite{sethia2013amplitude}, where the phase and amplitude are correlated.\\

To identify the coherent and incoherent domains in frequency, the frequency of the $i^{th}$ oscillator, $f_{i}$, is calculated from the time interval between adjacent zero crossings averaged over a finite number of crossings using the time series of its $x$ variable (after removing transients)\cite{gupta2019role}:
\begin{equation}
f_{i} = \frac{1}{L}\sum_{j=1}^{L}\frac{1}{t^{i}_{j+1}-t^{i}_{j}}
\end{equation}
where L is the number of zero crossings in one direction, and $t^{i}_{j}$ is the $j^{th}$ zero crossing point for $x_{i}$.\\

\begin{figure}
    \centering
    \includegraphics[width=0.5\textwidth]{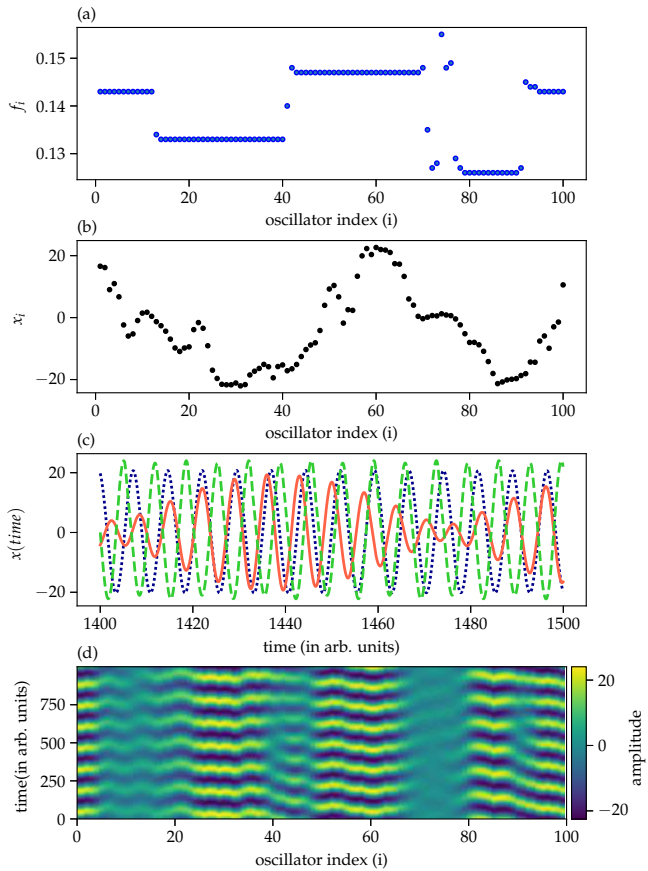}
    \caption{Frequency chimera state for 100 locally coupled R\"{o}ssler oscillators with coupling strength ($\varepsilon$) = 0.7, mismatch parameter ($\tau$) = 0.76, and number of slow oscillators, $N_s$ = 50. (a) Frequencies ($f_{i}$), (b) snapshots ($x_{i}$) at the last instant of time, and (c) time series from three different nodes 25 (blue, dotted) and 60 (green, dashed) from the coherent domain, and 41 (red, solid) from the incoherent domain. (d) Spatiotemporal plot for the emergent dynamics. The slow oscillators are randomly distributed and the network is initialized with random values in [-1,1].}
    \label{fig:ross_freq_chimera}
\end{figure}

In Fig. \ref{fig:ross_freq_chimera} we show the frequency chimera state with coexisting domains of coherence and incoherence in frequency for $\varepsilon$ = 0.7, and $\tau$ = 0.76. The frequency of the oscillator at every node and the snapshot of its $x$ variable, are shown in Figs. \ref{fig:ross_freq_chimera} (a) and (b), respectively. By studying the Fourier transform and time series of a few typical oscillators from coherent and incoherent domains, we find the nodes in coherent domains self-organize to an emergent frequency of periodic oscillations while those in incoherent regions, are in two frequency states. For lower values of coupling strength, the incoherent regions can be in chaotic states. The time series from three typical nodes from coherent and incoherent domains are shown in Fig. \ref{fig:ross_freq_chimera} (c), and the spatiotemporal behavior in Fig. \ref{fig:ross_freq_chimera} (d).\\

\begin{figure}
    \centering
    \includegraphics[width=0.45\textwidth]{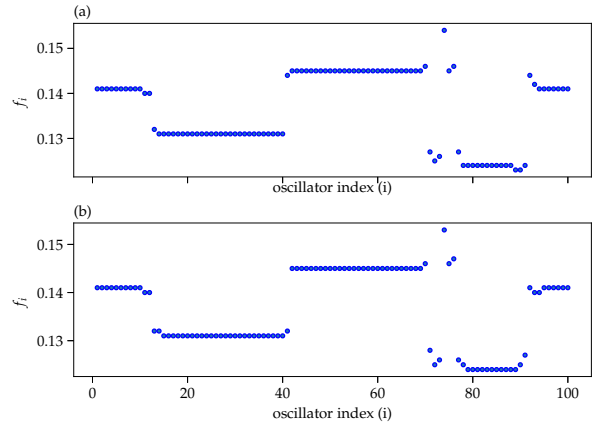}
    \caption{Frequency chimera state at two different instants of time. Frequencies calculated after (a) 50000, and (b) 100000 time steps of size 0.05 (after discarding transients). This indicates that the frequency chimera state is stable and persists for a long time.}
    \label{fig:stability}
\end{figure}

We evolve the system for longer times and observe the same behavior where a few oscillators are in incoherent frequencies while most of them settle to clusters of two or more coherent frequencies (Fig. \ref{fig:stability}). When repeated for a large set of different initial conditions, with the same distribution of slow nodes we find that the frequency chimera state prevails for all of them for the same parameters as shown in Fig. \ref{fig:ross_freq_chimera}. However, the location of clusters and their frequencies may change for a different distribution of slow and fast nodes in the network. When a small random perturbation of strength $\Delta \leq 0.5 $  is added to every node, we find the system settles back to the same frequency chimera state.

\section{Chimeralike states} \label{sect:chimeralike}

\begin{figure}
    \centering
    \includegraphics[width=0.5\textwidth]{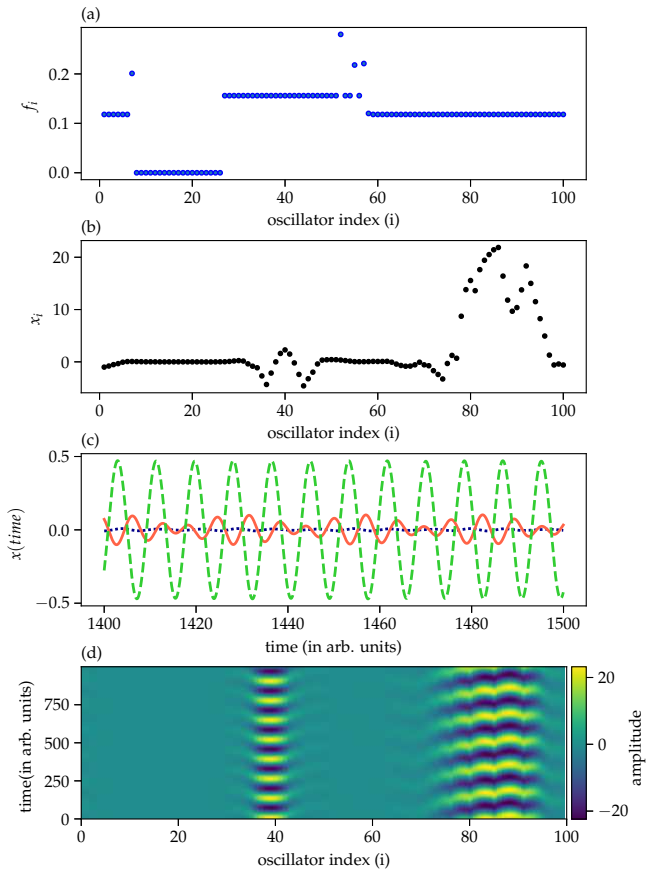}
    \caption{Chimeralike state for 100 locally coupled R\"{o}ssler oscillators. Here coupling strength ($\varepsilon$) = 0.7, mismatch parameter ($\tau$) = 0.7, and $N_s$ = 50. (a) Frequencies ($f_{i}$), (b) snapshots ($x_{i}$) at the last instant of time. (c) Time series from three different nodes 13 (blue, dotted) from AD, and 56 (red, solid) and 63 (green, dashed) from the incoherent domain. (d) Spatiotemporal plot.}
    \label{fig:ross_chimera_like}
\end{figure}

As $\tau$ decreases for the same coupling strength, the system self-organizes into two types of spatial clusters. We see partial amplitude death (AD) with some of the oscillators in AD while the others form clusters with incoherent oscillatory states. These types of states have been reported in the literature as chimeralike states \cite{dutta2015spatial}. A typical case is shown in Fig. \ref{fig:ross_chimera_like} where the oscillating clusters have either settled to an emergent frequency or are in two frequency state while others have settled to AD.\\

\begin{figure}
    \centering
    \includegraphics[width=0.5\textwidth]{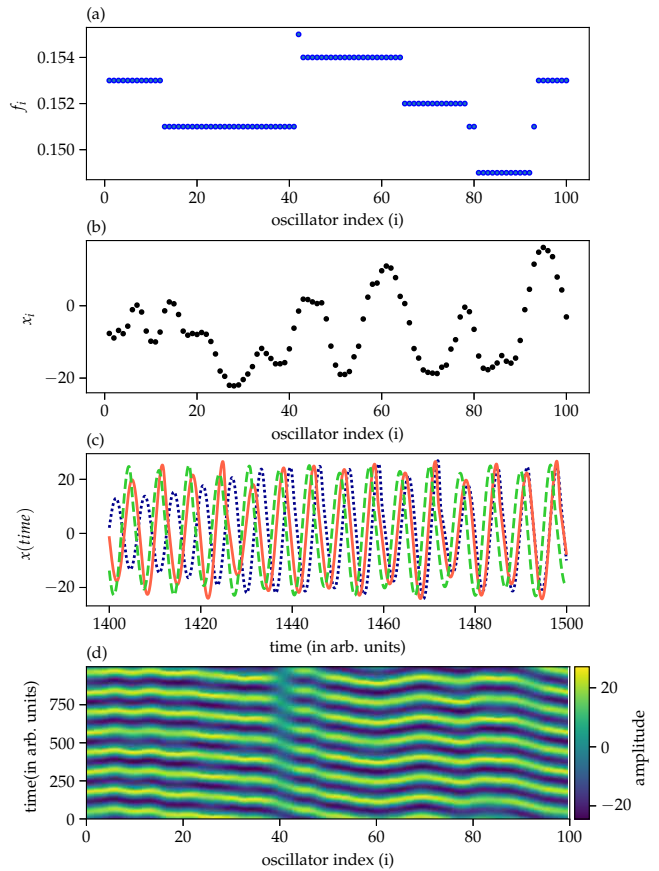}
    \caption{Chaotic traveling wave state or moving multicluster state for 100 locally coupled R\"{o}ssler oscillators. Coupling strength ($\varepsilon$) = 0.7, mismatch parameter ($\tau$) = 0.9, and number of slow nodes, $N_s$ = 50. (a) Frequencies ($f_{i}$), (b) snapshots ($x_{i}$) at the last instant of time, and (c) time series from three different nodes 10 (blue, dotted), 20 (red, solid) and 30 (green, dashed). (d) Spatiotemporal plot.}
    \label{fig:ross_CTW}
\end{figure}

As $\tau$ further decreases below 0.5, we find the whole network settles to a state of AD. Also with $\tau$ higher than 0.8, we observe chaotic traveling waves as shown in Fig. \ref{fig:ross_CTW}. When characterized with frequency, this state has multiclusters of differing frequency with the number and size of clusters changing in time, since the dynamics at each site is chaotic. With a further increase in $\tau$, the network reorganizes to a state of uniform emergent frequency or frequency synchronization.\\

\section{Characterization of frequency chimera}\label{sect:characterization}
In the system under study,  with random initial distribution of slow and fast oscillators, spontaneous clusters of oscillators can form with coherence in frequency due to their mutual interaction.  If the mismatch in timescales is less, these clusters can synchronize on a global scale, giving rise to frequency synchronized state($\tau > 0.9$), with an emergent frequency that is different from those corresponding to their intrinsic timescales. However, the mutual interaction of neighboring nodes competes with the intrinsic dynamical timescale that has a mismatch and as a result the oscillator can either settle into a coherent cluster, or remain incoherent. Since the initial distribution of slow and fast oscillators is random, different spontaneous clusters of frequency emerge in different locations.  With decrease in $\tau$ we observe them separated by incoherent oscillators that fail to synchronize with their neighbors, giving rise to frequency chimera state.
In addition, the interaction among slow and fast oscillators also results in amplitude suppression. For $\tau < 0.7$, the suppression of amplitude can drive some oscillators to amplitude death. It then manifests as chimeralike state in which some oscillators decay to a single fixed point while the rest can settle either to frequency clusters or remain chaotic. As $\tau$ decreases, ($\tau< 0.5$), the whole system goes to amplitude death (AD). Thus a random mixing of slow and fast oscillators with intermediate values of mismatch can display a variety of collective behaviors, such as multi-cluster states, frequency chimera states, and chimeralike states.\\

Since the coherence and incoherence are assigned using the frequencies, to understand the mechanism of onset of different emergent dynamics we compute the distribution of difference in frequencies, $z_i$, calculated as:
\begin{equation}
    z_i = \left | f_i - f_{i+1} \right |
\end{equation}

\begin{figure}
    \centering
    \includegraphics[width=0.5\textwidth]{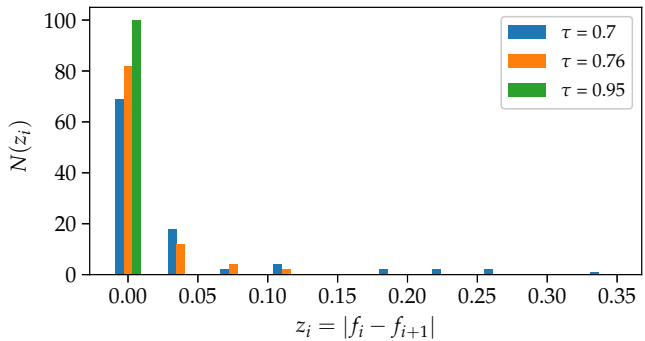}
    \caption{Distribution of $z_i$ for $\tau$ = 0.7 (chimeralike), 0.76 (frequency chimera), and 0.95 (frequency synchronization) for 100 locally coupled R\"{o}ssler oscillators with $N_s$ = 50 and $\varepsilon$ = 0.7. Slow and fast nodes are distributed randomly in the network and are initialized with random initial conditions in [-1,1]. The peaks are stacked near each other for clarity.}
    \label{fig:N_z}
\end{figure}

If all the oscillators are coherent in frequency, $z_i$ is 0 for all $i$, but nonzero for some $i$ if a few oscillators are incoherent. Thus this quantity will also give the extent of incoherence and coherence among the oscillators. This measure is usually employed for quantifying coherence in terms of amplitudes, but we modify it for frequencies in the present context\cite{chandran2019chimera,gopal2014observation}.\\
We plot the distribution, $N(z_i)$ for different parameter values in Fig. \ref{fig:N_z}. We see that for a large $\tau$ in timescale (for $\tau > 0.9 $), the oscillators are in frequency synchronized state within a single cluster, so there is a single peak at $z_i = 0$. As we decrease $\tau$, the oscillators settle to frequency chimera and for lower values of $\tau$, some of the oscillators reach the AD state, in isolated clusters, giving rise to a chimeralike state. In both these cases, $z_i$ is distributed over a range of values, the peak at 0 still giving coherent clusters of frequency synchronized or death states. On further decreasing $\tau$, all oscillators settle to a state of suppressed oscillations (AD) with $z_i$ peaking again at zero. 

The frequently used measure of chimera, strength of incoherence (S) \cite{gopal2014observation}, is adapted to characterize the new frequency chimera state by using frequencies of each node in its calculation. The first step is to divide the oscillators into bins of equal size $n = N/M$, and then calculate the local standard deviation in frequencies ($\sigma_{l}$) for bins $l = 1,2,...M.$ as follows:
\begin{equation}
\sigma_{l}(m)= \sqrt{\frac{1}{n}\sum_{j=n(m-1)+1)}^{mn}[f_{l,j}-\left \langle f_{l} \right \rangle]^{2}}
\end{equation}
where $f_{i}$ is the frequency of $i^{th}$ oscillator.\\ %check again

From $\sigma_{l}$, we can calculate strength of incoherence in frequencies, S(f) as:
\begin{equation}
S(f) = 1- \frac{\sum _{m =1}^{M}f_{m}}{M}, f_{m} = \Theta (\delta-\sigma_{l}(m))
\end{equation}

where $\Theta$ represents the Heaviside step function. $\delta$ is the threshold that decides coherence. Here, we set it to be 0.05 of the standard deviation. Then S(f) = 1 indicates incoherent state, S(f) = 0
indicates either a coherent state or amplitude death, and 0 $<$ S(f) $<$ 1 indicates a frequency chimera, chimeralike state, or multicluster state. \cite{gopal2014observation}.\\
\begin{figure}
    \centering
    \includegraphics[width=0.5\textwidth]{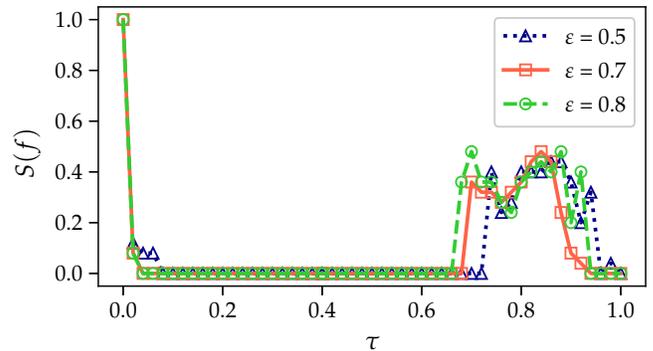}
    \caption{Strength of incoherence in frequencies, S(f) as a function of mismatch parameter, $\tau$, for 100 coupled R\"{o}ssler oscillators for coupling strengths $\varepsilon$ = 0.5 (blue, dotted), 0.7 (red, solid) and 0.8 (green, dashed). Number of slow oscillators, $N_s$ = 50. S(f) = 0 for amplitude death (AD) state and fully coherent state. For incoherent state, S(f) = 1, while it stays between 0 and 1 for frequency chimera, chimeralike, and multicluster states.}
    \label{fig:si_vs_e}
\end{figure}
The strength of incoherence $S(f)$ computed as above for the system under study is shown in Fig. \ref{fig:si_vs_e}. We see $S(f)$ = 0 for $\tau <
0.58$ and $\tau > 0.98$, with AD on one end and frequency synchronization on the other end over the full network. For very small values of $\tau$ we get incoherence with $S(f) = 1$. In between we get frequency chimera, chimeralike states, and chaotic traveling wave (or moving multicluster) states, with $0 < S(f) < 1$. Specifically, the values of $S(f)$ for the frequency
chimera state shown in Fig. \ref{fig:ross_freq_chimera} is 0.28, that for chimeralike state in Fig. \ref{fig:ross_chimera_like} is 0.35, and for a multicluster state or traveling chaotic waves in Fig. \ref{fig:ross_CTW} it is 0.08. Thus, for R\"{o}ssler dynamics studied here, with $\tau$ decreasing from 1 to 0 (with coupling strength at 0.7), the system undergoes transitions from frequency synchronization ($1 < \tau < 0.9$), to moving multicluster or chaotic traveling wave state ($0.8< \tau < 0.9$), to frequency chimera state ($0.8< \tau < 0.7$).  Upon further decreasing $\tau$, some nodes go to amplitude death, resulting in a chimeralike state ($0.7 < \tau < 0.5$). With lower values (i.e. $\tau < 0.5$), all nodes go to the amplitude death state. Finally, at very low values, there is no adjustment of frequencies and all the nodes show incoherence.\\

\subsection{Conditions for emergence of frequency chimera}
\begin{figure}
    \centering
    \includegraphics[width=0.5\textwidth]{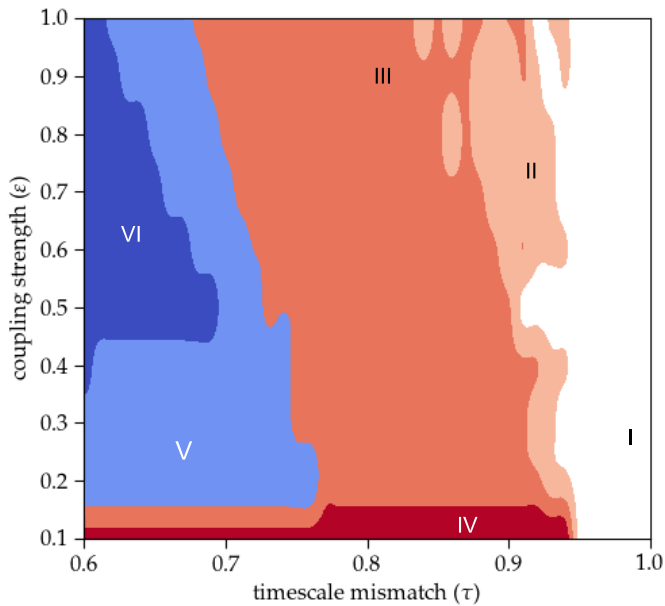}
    \caption{Parameter plane, ($\tau, \varepsilon$) for 100 locally coupled R\"{o}ssler oscillators  with 50 slow nodes distributed randomly. Region I: frequency synchronization, II: chaotic traveling wave or multicluster state, III: frequency chimera, IV: incoherence, V: chimeralike state, and VI: amplitude death (AD) state.}
    \label{fig:para_plane_ross_freq}
\end{figure}

The emergence of different types of spatiotemporal dynamics presented above mostly depends on the parameters $\tau$ and coupling strength $\varepsilon$. We plot this parameter plane to indicate the regions where these different emergent dynamics can occur in Fig.
\ref{fig:para_plane_ross_freq}. The regions of various dynamical states are identified by using the characterizing measure, $S(f)$ and average amplitude of oscillators. Region I corresponds to the frequency synchronized state, with $S(f) = 0$. Region II indicates moving multicluster or chaotic travelling wave state [$S(f) < 0.2$], that occurs as $\tau$ decreases for a given $\varepsilon$. The frequency chimera state exists in the region III [$0 <S(f) <1$], while region IV corresponds to the incoherent state [$S(f) = 1$]. The chimeralike [$0 <S(f) < 1$, with partial amplitude death) and AD ($S(f)=0$, with all nodes in AD] states exist in regions V and VI respectively. We note the frequency chimera state prevails for a wide range of the system parameters $\tau$ and $\varepsilon$.\\

\begin{figure}
    \centering
    \includegraphics[width=0.5\textwidth]{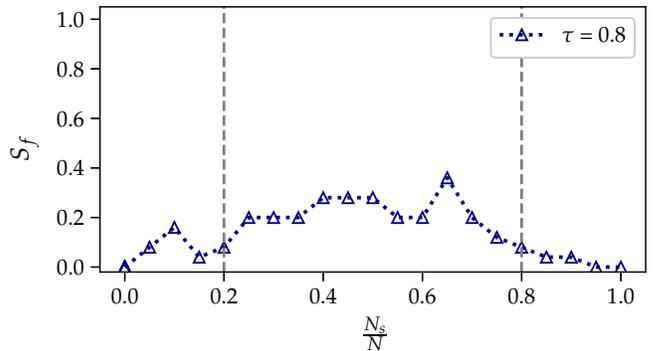}
    \caption{Variation of strength of incoherence with number of slow nodes, for $\varepsilon = 0.7$, $\tau = 0.8$. The frequency chimera for this particular set of parameters occurs for $0.2<\frac{N_{s}}{N}<0.8$, as indicated by the region within dashed lines (identified by existence of incoherent nodes).}
    \label{fig:var_slow}
\end{figure}

In addition to the above parameters, and their range of values for the emergence of frequency chimera and other states mentioned above, we find that the onset of chimera depends critically on the number of slow nodes. For this, we compute the $S(f) $ values for $\varepsilon = 0.7$, $\tau = 0.8$ and varying the number of slow nodes in the system (Fig. \ref{fig:var_slow}). Corresponding to all slow and all fast cases, at the extreme ends, we see coherence in frequency while in between for a range of values of the fraction of slow systems ($\frac{N_{s}}{N}$), frequency chimera states emerge. Here, we distinguish frequency chimera states from multicluster states by the existence of incoherent nodes in the former, indicated by vertical grey lines in the Fig. \ref{fig:var_slow}. This range decreases with an increase in $\tau$.
Thus for a given $\tau$ and $\varepsilon$ within the region III of Fig. \ref{fig:para_plane_ross_freq},  we require a specific range of values for $\frac{N_{s}}{N}$ for frequency chimera to emerge. In Fig. \ref{fig:var_slow}, frequency chimera occurs for the range $0.2< \frac{N_{s}}{N}< 0.8$. On both sides, for $0< \frac{N_{s}}{N}< 0.2$ and $0.8< \frac{N_{s}}{N}< 1$, we find multicluster oscillations.

\begin{figure}
    \centering
    \includegraphics[width=0.5\textwidth]{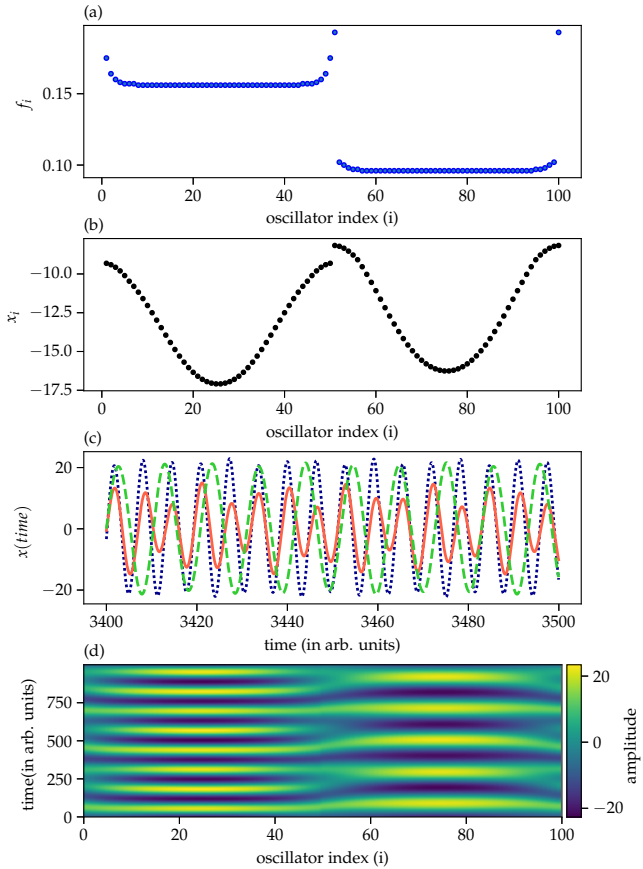}
    \caption{Chimera state for 100 nonlocally coupled (P = 5) R\"{o}ssler oscillators, where first 50 are slow  with $\tau$ = 0.6 and the next 50 are fast. In this case, we observe two coherent clusters of frequency (with phase shift within each cluster), and some incoherent oscillators at the boundary of these two regions. Coupling strength, $\varepsilon$ = 0.6. (a) Frequencies ($f_{i}$), (b) snapshots ($x_{i}$) at the last instant of time, and (c) time series from three different nodes 20(blue, dotted), 45 (red, solid) and 70(green, dashed). (d) Spatiotemporal plot for the emergent dynamics.}
    \label{fig:nonlocal_first_50_chimera}
\end{figure}
We extend the study to another specific configuration where the first 50 are fast and next 50 are slow in their timescales. We find that the system settles to two clusters of differing frequencies with incoherent oscillators existing at the boundary.
With increase in the range of coupling or nonlocal coupling, we get chimera states in which two clusters of frequency synchronized state coexist, with oscillators within each cluster shifted in phase and incoherent systems at the boundary, as indicated in Fig. \ref{fig:nonlocal_first_50_chimera}.\\

We also study the case where periodic R\"{o}ssler oscillators form the intrinsic dynamics and find that with the mismatch in dynamical timescales, mostly chimeralike state prevails over a range of coupling strength.
% \textcolor{red}{Interestingly, we find frequency chimera states here as well, with the incoherent oscillators settling to a two-frequency state.}\\

\section{frequency chimera states in coupled van der Pol systems}\label{sect:freq. chimera vdP}
\begin{figure}
    \centering
    \includegraphics[width=0.5\textwidth]{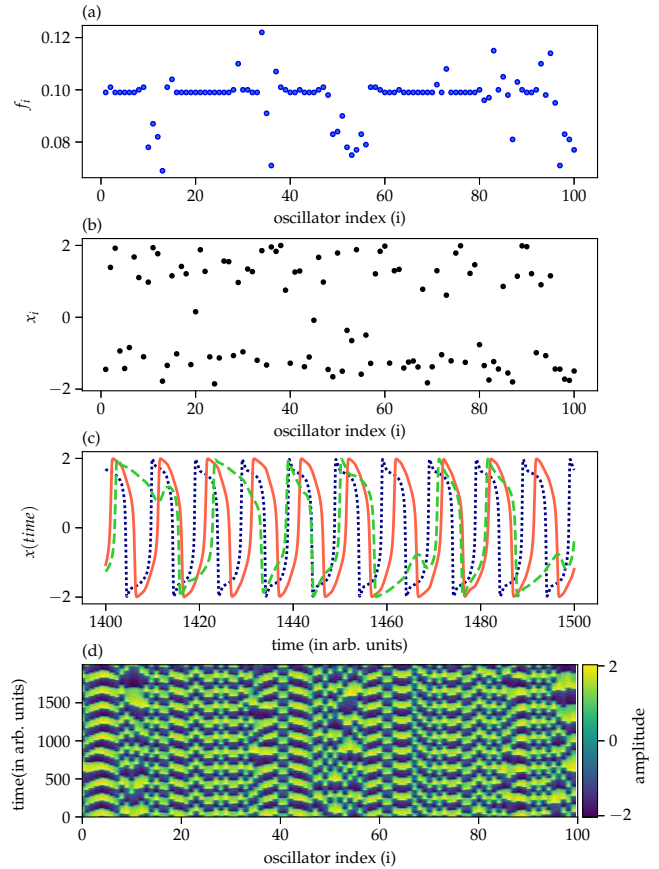}
    \caption{Frequency chimera state exhibited by locally coupled (forced) van der Pol oscillators, at $\varepsilon$ = 0.1, $\tau$ = 0.5, $N_s$ = 50. (a) Frequencies ($f_{i}$), (b) snapshots ($x_{i}$) at the last instant of time, (c) time series from nodes 20 (blue, dotted), and 70 (green, dashed) from the coherent region, and node 34 (red, solid) from the incoherent region. (d) Spatiotemporal plot.}
    \label{fig:freq_chimera_vdP}
\end{figure}

In this section, we present the existence of similar emergent states observed in a ring of coupled driven van der Pol oscillators, given in eq. (\ref{eq:vdp}). With local coupling and random initial conditions, we find frequency chimera state as shown in Fig. \ref{fig:freq_chimera_vdP}. However in this case, the different coherent clusters settle to a single frequency, unlike the multichimera state exhibited by coupled R\"{o}ssler oscillators of section \ref{sect:freq_chimera}. The coherent clusters display a periodic behavior, while the incoherent ones display chaos.\\
With intrinsic dynamics in the periodic regime ($A = 0$), for nonlocal coupling (P = 10), we find frequency chimera states. We also observe chimeralike states with coexistence of incoherent oscillators and coherent clusters with suppression of dynamics. However, these states display oscillation death state for the coherent clusters, which is different from AD observed in the case of chimeralike states in coupled R\"{o}ssler oscillators.\\

\section{Conclusion}\label{sect:conclusion}
The emergence of chimera states in coupled oscillators has been associated with
heterogeneity in otherwise identical systems. Such heterogeneity could be introduced in multiple
ways, such as network topology, range of coupling, delay, initial conditions, etc. In many real
systems, heterogeneity can arise due to differences in timescales of evolution. In such cases, a
simple network of coupled systems with different dynamical timescales can serve as a model to
study their emergent dynamics. The possibility of the occurrence of chimera states in such systems is a question of relevance due to the ubiquity of real systems displaying heterogeneity in timescales.\\

In this article we report the study of a type of chimera called frequency chimera that
can exist in timescale mismatched systems, with only local coupling and random initial
conditions. The study includes two types of nonlinear oscillators, R\"{o}ssler and forced van der Pol systems. Underlying the apparent incoherence in spatiotemporal patterns, we find there exist domains with a certain level of coherence in terms of the frequency of the oscillators. In this context, to characterize the different possible states of spatiotemporal dynamics, we redefine
the measure of strength of incoherence, in terms of frequencies and compute its values for a
range of parameter values. We also find chimeralike state with part of the system in amplitude
death and multiclusters of frequency synchronization on varying the parameters. We present the $\tau$-$\varepsilon$ parameter plane that indicates the prevalence of frequency chimera and the regions of different possible emergent states.\\
Thus a random distribution of slow and fast oscillators, coupled with sufficiently strong mismatch in timescales and coupling strength facilitates the emergence of frequency chimera state. It is the intermediate state of the system, between amplitude death and frequency synchronization. The required heterogeneity in this case is the mismatch in dynamical timescales among otherwise identical oscillators. We note frequency chimera state due to nonisochronicity parameter was reported earlier in a different context\cite{premalatha2015impact}.\\
Our study reveals how the presence of differing dynamical timescales leads to
the organization of coupled systems into frequency chimeras, where the coexisting coherence and incoherence are defined with respect to the frequency of their oscillations. They also can have other emergent states like chimeralike states with coexisting amplitude death and oscillatory states or settle to a common emergent
frequency. The emergence of such diverse states in complex systems is not uncommon and our study reveals yet another model to understand their dynamics. The cognitive activity of the brain depends on the emergence of spatiotemporal
patterns emerging from collective regional or local activities and hence a variety of states of partial synchrony can form, including chimera states \cite{Bansaleaau8535}. We propose that such patterns can arise from the differing timescales at the neuronal level where the difference in dynamical timescales can be attributed to the
differences in the properties of the axonal membranes in individual biological neurons. Hence
our study has relevance in understanding brain functions involving chimera states. Another real-world system of practical importance is power grids and transmission networks, which are susceptible to chimera states, where the level of coherence expected is mostly in the frequency of the states\cite{panaggio2015chimera}. A recent study also indicates that frequency clusters that occur in coupled systems are very sensitive to external signals and such states of partial synchronization are utilized by natural systems to optimize their ability to detect weak signals \cite{faber2021chimera}. Since mismatch in dynamical timescales can occur in many
real-world systems, we hope the present study is relevant to understand their complex spatiotemporal dynamics.\\

\section{acknowledgments}
SK acknowledges financial support from the Council of Scientific and Industrial Research (CSIR), Government of India.

%\bibliography{refs}
\end{document}